\documentclass[12pt]{article}
\usepackage{emulateapj}
\usepackage{apjfonts}
\usepackage{flushrt}
\usepackage{epsfig}
\submitted{Submitted to the Astrophysical Journal, 3 March 1999}
\newcommand{\etal} {{\it et~al.\ }}
\newcommand{\dm} {\,{\rm cm}^{-3}\,{\rm pc}}
\newcommand{\psr}{PSR\,B1620$-$26}
\def\spose#1{\hbox to 0pt{#1\hss}}
\newcommand\lsim{\mathrel{\spose{\lower 3pt\hbox{$\mathchar"218$}}
     \raise 2.0pt\hbox{$\mathchar"13C$}}}
\newcommand\gsim{\mathrel{\spose{\lower 3pt\hbox{$\mathchar"218$}}
     \raise 2.0pt\hbox{$\mathchar"13E$}}}
\begin{document}
\title{The Triple Pulsar System \psr\ in M4}

\author{S. E. Thorsett\altaffilmark{1}}
\altaffiltext{1}{Alfred P. Sloan Research Fellow}
\affil{Joseph Henry Laboratories and Department of Physics, Princeton
  University, \\Princeton, NJ 08544; steve@pulsar.princeton.edu}

\and
\author{Z. Arzoumanian}
\affil{Department of Astronomy, Cornell University, Ithaca, NY 14853}

\and
\author{F. Camilo\altaffilmark{2} and A. G. Lyne}
\altaffiltext{2}{Marie Curie Research Fellow}
\affil{The University of Manchester, Nuffield Radio Astronomy
  Laboratories, Jodrell Bank, Macclesfield, Cheshire SK11\,9DL UK}

\begin{abstract}
  The millisecond pulsar \psr, in the globular cluster M4, has a white
  dwarf companion in a half-year orbit. Anomalously large variations
  in the pulsar's apparent spin-down rate have suggested the presence
  of a second companion in a much wider orbit. Using timing
  observations made on more than seven hundred days spanning eleven
  years, we confirm this anomalous timing behavior.  We explicitly
  demonstrate, for the first time, that a timing model consisting of
  the sum of two non-interacting Keplerian orbits can account for the
  observed signal.  Both circular and elliptical orbits are
  allowed, although highly eccentric orbits require improbable orbital
  geometries.
  
  The motion of the pulsar in the inner orbit is very nearly a
  Keplerian ellipse, but the tidal effects of the outer companion
  cause variations in the orbital elements.  We have measured the
  change in the projected semi-major axis of the orbit, which is
  dominated by precession-driven changes in the orbital inclination.
  This measurement, along with limits on the rate of change of other
  orbital elements, can be used to significantly restrict the
  properties of the outer orbit.  We find that the second companion
  most likely has a mass $m\sim0.01M_\odot$--- it is almost certainly
  below the hydrogen burning limit ($m<0.036M_\odot$, 95\%
  confidence)---and has a current distance from the binary of
  $\sim35$\,AU and orbital period of order one hundred years.
  Circular (and near-circular) orbits are allowed only if the pulsar
  magnetic field is $\sim3\times10^9$\,G, an order of magnitude higher
  than a typical millisecond pulsar field strength. In this case, the
  companion has mass $m\sim1.2\times10^{-3}M_\odot$ and orbital period
  $\sim62$\,years.
\end{abstract}

\section{Introduction}
It is now a quarter century since the discovery of the first binary
radio pulsar.  Since 1974, about fifty other binaries have been
discovered, including systems with a wide variety of companions:
planets, ``brown dwarfs'' below the hydrogen burning limit, white
dwarfs, neutron stars, and massive main sequence stars.

In 1993, the first identification of a pulsar in a triple system was
proposed. \psr\ is an 11\,ms pulsar associated with the nearby
globular cluster M4.  It is in a 191\,day, low-eccentricity orbit with
a $\sim0.3M_\odot$ companion presumed to be a white dwarf
(\cite{lbb+88,ml88}). However, the early pulse timing data were not
well described by a simple Keplerian model (\cite{tho91b}).  Backer
(1993)\nocite{bac93} was the first to realize that the data could be
well fitted with the addition of a substantial cubic term to the
pulsar timing model: a change in the apparent spin-down rate of the
pulsar that was so large that it predicted a change in sign of the
pulsar frequency derivative---from spin-down to spin-up---on a
timescale of $\sim10$\,yrs. Such a cubic could be induced by a
changing acceleration of the pulsar binary under an external
gravitational force, which Backer proposed could be due to a second
companion in a wide orbit. Further timing observations
(\cite{tat93,bfs93,bt95,ajrt96}) led to measurements of the next two
terms in the Taylor expansion of the pulsar frequency as well as to
the first measurements of perturbations in the orbital elements of the
inner binary caused by the tidal effects of the outer body.  These
observations in turn led to tighter constraints on the properties of
the second orbit; the second companion most likely has a mass typical
of a brown dwarf or planet, $\sim0.01M_\odot$, and is in a
$\sim40$\,AU orbit (\cite{ras94,sig95,ajrt96,jr97}).

We have now been observing \psr\ for over a decade.  In
\S\ref{sec:obs}, we describe the various observing systems that have
been used.  In \S\ref{sec:model}, we describe the analysis of the
data.  For the first time, we present results from a full, two-orbit
analysis, rather than a single Keplerian orbit combined with a
Taylor series expansion in pulsar frequency. We describe the
constraints that can be placed on the elements of both the inner and
outer orbits from the timing data.  Finally, in
\S\ref{sec:discussion}, we discuss the implications of our
measurements for the understanding of the triple
system. 

\section{Observations and Analysis Summary}\label{sec:obs}

Since shortly after its discovery, \psr\ has been observed regularly
at multiple radio frequencies using several different telescopes.  We
report on observations made using the Very Large Array (VLA), near
Socorro, New Mexico; the 43\,m telescope at the National Radio
Astronomy Observatory in Green Bank, West Virginia; and the 76\,m Lovell
Telescope at Jodrell Bank, England.

At the VLA, observations were made on one or two days each two or
three months from 1990 November 30 to 1998 September 21, excluding the
period from 28 June 1996 to 28 September 1997---a total of 49 days.  A
filter bank and the VLA's ``High Time-Resolution Processor'' were used
to divide a 50\,MHz bandpass at 1.66\,GHz into 14 slightly overlapping
4\,MHz channels in each of two orthogonal circular polarizations.  The
signal in each frequency channel was sampled and averaged
synchronously with the predicted topocentric pulsar period, then the
channels were shifted to account for dispersion and added, and the
polarizations were summed, to produce a single integrated pulse profile
every 5\,minutes, using a Princeton Mark\,III pulsar timing system
(\cite{skn+92}). The start time of each integration was referenced to
external time standards using GPS. After eliminating data contaminated
with radio-frequency interference (RFI), 486 individual arrival time
measurements were available, comprising just over 40\,hrs of observing
time. Timing precision varied, with a weighted, root-mean-square precision
of 31\,$\mu$s.

At Green Bank, observations were made in campaigns that lasted several
days each quarter from 1989 August 20 to 1998 August 7---a total of
165 days.  At each epoch observations were made at two frequencies,
varying between 400, 575, 800, and 1330\,MHz.  The ``Spectral
Processor'' fast-fourier transform spectrometer was used to synthesize
512 channels across a 40\,MHz passband (256 channels across 20\,MHz
before 1991 February) in each of two orthogonal polarizations, and to
fold the channels synchronously with the topocentric pulsar period.
The channels and polarizations were summed, as at the VLA, to produce
a single integrated pulse profile for each integration period.
Each integration was begun on a time signal from the site maser, and
GPS was used to reference observatory time to external standards.
After eliminating data contaminated with RFI, 763 arrival time
measurements were available, comprising just over 60\,hrs of observing
time, with a weighted, r.m.s. precision of 40\,$\mu$s.

At Jodrell Bank, observations were made on 490 days between 1987
October 15 and 21 October 1998.  On each day, observations were made
at either 400, 600, 1400, or 1600\,MHz. At each frequency, two
circular polarizations were observed, using a $2\times64\times0.125$\,MHz
filterbank at 400 and 600\,MHz, and a $2\times32\times1$\,MHz 
filterbank at higher frequencies. The signals were detected, filtered, 
and the polarizations added, then were sampled and averaged
synchronously with the predicted topocentric pulsar period. Shifting
and adding of channels to account for interstellar dispersion was done 
in hardware.  There were 537 arrival time measurements available, with
weighted, r.m.s. precision of 45\,$\mu$s.

In total, the data set spans exactly 31,391,908,721 pulsar rotations.

Analysis of the pulse arrival times was carried out with the standard
software package {\sc Tempo} (\cite{tw89}).  The integrated profiles
were cross-correlated with a high-signal-to-noise-ratio average
profile to measure the offsets between the start of each integration
and the arrival time of a pulse near the center of the integration
(\cite{tay92}). The arrival times were fitted with a model that
included the pulsar phase at a reference epoch, the pulsar frequency
$f$ and frequency derivative $\dot f$, the pulsar position ($\alpha$,
$\delta$) and proper motion ($\dot\alpha\cos\delta$,$\dot\delta$), the
dispersion measure DM and a linear rate of change of the dispersion
measure, and the parameters of the known binary orbit: the period
$P_a$, projected semi-major axis $x_{1a}=a_{1a}\sin i_a/c$,
eccentricity $e_a$, argument of periastron $\omega_{1a}$ (measured
from the ascending node to periastron in the orbital plane), and a time of
periastron $T_a$, as well as other parameters discussed
below.\footnote{Note that to simplify later discussion, we use the
  suffixes $a$ and $b$ to distinguish the two Keplerian orbits that
  will be discussed, and $1$ and $2$ in each case to distinguish the
  pulsar and companion. This allows us to unambiguously distinguish,
  for example, the semi-major axis of the pulsar and companion in the
  first orbit---$a_{1a}$ and $a_{2a}$, respectively---as well as the
  semi-major axis of the pulsar in its first and second
  orbits---$a_{1a}$ and $a_{1b}$, respectively. Subscripts will be
  dropped where redundant; for example, the orbital period in the
  first orbit is $P_a\equiv P_{1a}\equiv P_{2a}$.}  The parameters
were varied, and the differences between the model and observed
arrival times were minimized in a least-squares sense. The data are
weighted to account for large variations in their signal-to-noise
ratios. Quoted uncertainties are, unless otherwise noted, intended to
be approximately 68\% confidence limits; in general, they are obtained
by testing the robustness of the fitted parameter estimates and formal
uncertainties when subsections of data are omitted, and through
bootstrap Monte Carlo techniques.  Typically, quoted error regions are
between one and three times as broad as the formal uncertainties
reported by the {\sc Tempo} fitting procedure.  Timing
``residuals''---the observed pulse arrival times minus the model
predictions---are calculated; examples of the observed
residuals will be shown below.

\section{Timing Results and Discussion}\label{sec:model}

In Fig.\,1, we display the timing residuals for a
model containing only a fit for $f$, $\dot f$, the astrometric
parameters, and a single Keplerian orbit.  Clearly, the model poorly
fits the data; the residuals are dominated by the cubic term reported
by Backer (1993). That the residuals are cubic in form---rather than
linear or quadratic---should not
surprise us, since the model that has already been removed from the data
contains a second order polynomial:
$\phi(t)=\phi_0+f(t-T_0)+\frac{1}{2}\dot f(t-T_0)^2$, where $T_0$ is an
arbitrary time chosen near the center of the data span.

\centerline{\epsfig{file=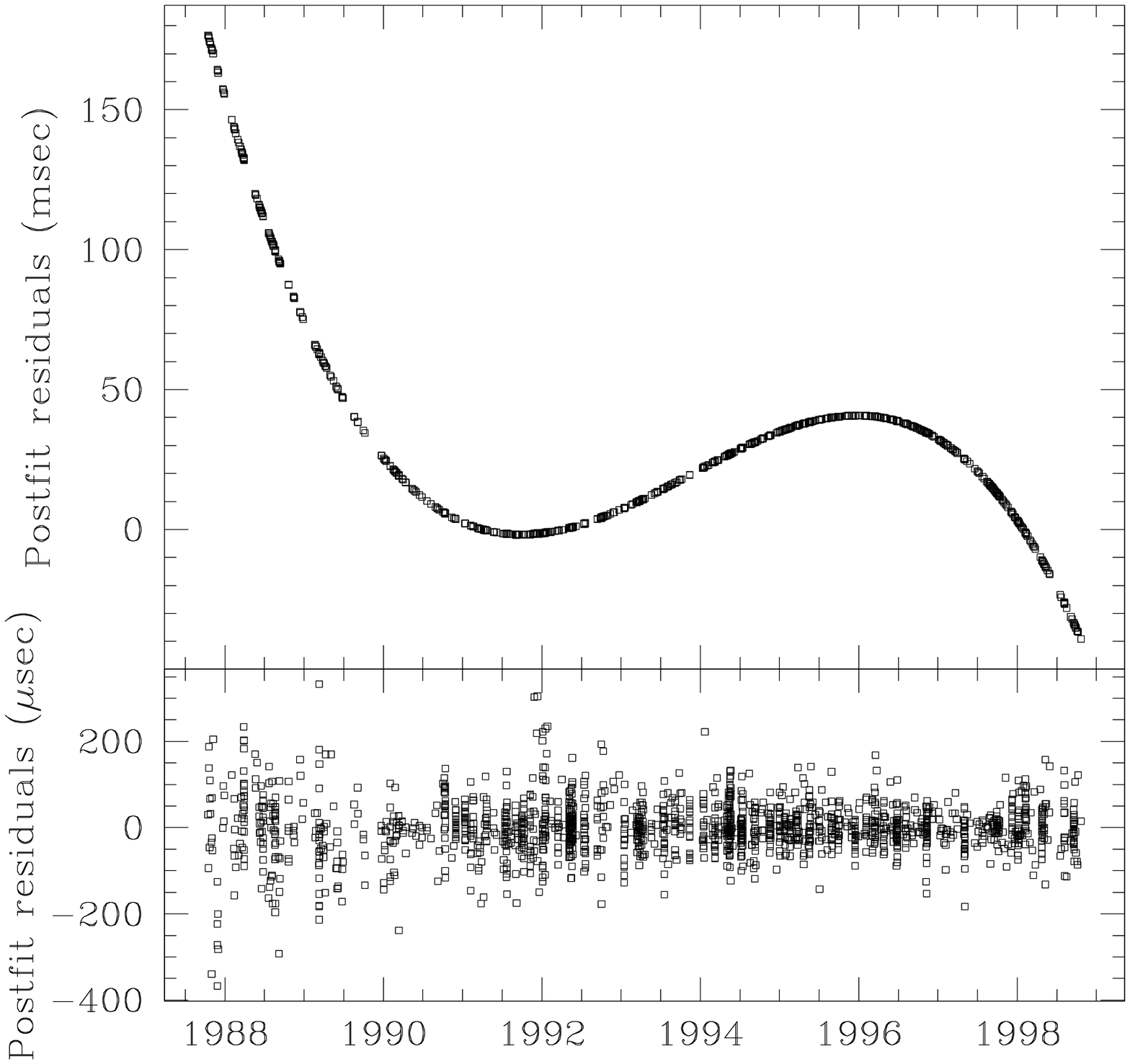,width=0.9\columnwidth}}
\figcaption{\label{fig:keponly}In the top panel, the best fit timing
  model containing only a fit for $f$, $\dot f$, the astrometric
parameters, and a single Keplerian orbit has been removed from the
data. A large cubic term dominates the postfit timing residuals.  In
the bottom panel, the timing model given in Table~1 has been removed
from the data.  Note the difference in scales between the panels.}

Qualitatively similar timing behavior is familiar from studies of
young pulsars (e.g., \cite{dr83}). So-called ``red'' timing noise is
thought to arise from stochastic interactions between the crust of the
neutron star and the superfluid vortices that carry angular momentum
in the interior of the star.  There are strong arguments against such
an interpretation in the case of \psr.  First, our understanding of
timing noise suggests that it is inversely correlated with the pulsar
age: old, relatively cold millisecond pulsars have relatively little
phase wander (\cite{srtr90,cor93,antt94}).  Second, the magnitude of the
observed cubic is very large for any pulsar, independent of age.
During the eleven years of observations, the pulsar's apparent
spin-down rate has varied from $-8.1\times10^{-15}\mbox{s}^{-2}$ to
$-1.4\times10^{-15}\mbox{s}^{-2}$, changing the apparent torque by a
factor of more than five; at the current rate of change, the
spin-down of the pulsar will change to an apparent spin-up in
November 2000.  While we cannot rule out intrinsic spin instabilities
as the cause of the observed signal, such instabilities are, in
magnitude, unlike any seen in any other pulsar, and have no known
theoretical basis.

Turning to external causes, the most promising explanation of the
large $\ddot f$ is a changing gravitational acceleration (``jerk'') by
a massive body external to the binary.  As we have previously
discussed (\cite{tat93}), the most likely candidate is a second
gravitationally-bound companion. Although several pulsars in globular
clusters are observed to have positive apparent frequency derivatives
due to acceleration in the mean cluster field
(\cite{wkm+89,and92,nt92}), the central mass density and velocity 
dispersion in M4 is far too low to produce the observed jerk on the
\psr\ system (\cite{phi93}).  A close passage by a cluster star on a
hyperbolic orbit cannot be ruled out, but the chance of discovering
the binary during such a transient event is very small
($\sim2\times10^{-5}$, \cite{phi93}).  Timing measurements could
eventually distinguish between ellipsoidal and hyperbolic orbits for a 
second companion, but we will not further consider the possibility here.

\subsection{A Polynomial Model}\label{sec:poly}
Obviously, a theory-independent way to characterize the deviations of
the observations from the standard timing model is to add additional
terms to the polynomial for $\phi(t)$, essentially Taylor-expanding
the residuals around the epoch $T_0$.  Hence $\phi(t)=\phi_0+f(t-T_0)+
\frac{1}{2}f^{(1)}(t-T_0)^2+\frac{1}{6}f^{(2)}(t-T_0)^3+\cdots$, where
we have introduced the convenient notation $f^{(n)}\equiv
d^n\!f/dt^n$. Given a model for the system, such as acceleration of
the binary in the gravitational field of a second companion, the
measured $f^{(n)}$ can be related to the model parameters.  This is the
approach that has previously been used in analysis of \psr\ data
(\cite{tat93,jr97}). 

The binary pulsar model can be further generalized to allow secular or
periodic variations in the orbital elements.  The simplest extension
is to allow the elements to vary linearly with time: for example,
$\omega=\omega_0+\dot\omega(t-T_0)$, where both $\omega_0$
and $\dot\omega$ are free parameters in the least-squares fit. In the
limit of a Keplerian orbit, of course, we expect the time derivatives
of the elements to vanish.

\begin{deluxetable}{ll}
\tablecolumns{2}
\tablewidth{0pc}
\tablecaption{Timing parameters of PSR\,B1620$-$26\label{tab:poly}}
\tablehead{\colhead{Parameter}&\colhead{Value (error)}}
\startdata
Right ascension (J2000.0) & $16^{\mbox{\scriptsize
h}}23^{\mbox{\scriptsize m}}38\fs2218(2)$ \nl
Declination (J2000.0) & $-26^\circ31'53\farcs769(2)$ \nl
Proper motion RA (mas\,yr$^{-1}$) & $-$13.4(1.0) \nl
Proper motion Dec (mas\,yr$^{-1}$) & $-$25(5) \nl
Dispersion measure ($\dm$) & 62.8633(5) \nl
$dDM/dt$ ($\dm$\,yr$^{-1}$)       & $-0.0005(2)$ \nl
\nl
Spin period $P$ (ms) & 11.0757509142025(18) \nl
Spin frequency $f$ (Hz) & 90.287332005426(14) \nl
$\dot f$ (s$^{-2}$) & $-5.4693(3)\times10^{-15}$ \nl
$\ddot f$ (s$^{-3}$) & $1.9283(14)\times10^{-23}$ \nl
$f^{(3)}$ (s$^{-4}$) & $6.39(25)\times10^{-33}$ \nl
$f^{(4)}$ (s$^{-5}$) & $-2.1(2)\times10^{-40}$ \nl
$f^{(5)}$ (s$^{-6}$) & $3(3)\times10^{-49}$ \nl
Epoch of $f$ (JD) & 2448725.5 \nl
\nl
Projected semi-major axis $x_{1a}$ (s) & 64.809460(4) \nl
Orbital period $P_a$ (days) & 191.44281(2) \nl
Eccentricity $e_a$ & 0.02531545(12) \nl
Time of periastron $T_a$ (JD) & 2448728.76242(12) \nl
Argument of periastron $\omega_{1a}$ $(^\circ)$ & 117.1291(2) \nl
Mass function ($M_\odot$) & $7.9748\times 10^{-3}$ \nl
\nl
$\dot x_{1a}$ & $-6.7(3)\times10^{-13}$ \nl
$\dot P_a$ & $4(6)\times10^{-10}$ \nl
$\dot e_a$ (s$^{-1}$) & $0.2(1.1)\times10^{-15}$ \nl
$\dot\omega_{1a}$ ($^\circ\mbox{yr}^{-1}$) & $-5(8)\times10^{-5}$ \nl
\enddata
\tablecomments{Timing parameters relative to a model including a
  Keplerian orbit with elements allowed to vary linearly with time,
  and a Taylor expansion in the barycentric pulsar frequency extended
  through $f^{(5)}$. Position is relative to the JPL~DE405 solar
  system ephemeris. Numbers in parentheses are uncertainties in the last
  digits shown. See \S\protect\ref{sec:obs} for discussion of
  notation.}
\end{deluxetable}

In Table\,\ref{tab:poly}, we present the timing parameters obtained
from a fit to the \psr\ data using a model including frequency
derivatives to $f^{(5)}$ and allowing for linear variations in the orbital
elements.
In general, the results are in excellent agreement with those we have
previously published, with substantially smaller uncertainties.  (Note
that, in order to reduce parameter covariances, the epoch of the
frequency expansion has been changed from that of Thorsett \etal\ 
(1993), to a point nearer the center of the current data set.)

A significant measurement of the pulsar proper motion has been made
for the first time; it can be compared with optical measurements of
the cluster proper motion (\cite{ch93b}): $\mu_{\rm RA}=-11.6(7)$ and
$\mu_{\rm dec}=-15.7(7)\mbox{mas\,yr}^{-1}$.  The measurements
disagree at $\sim95\%$ confidence---the difference is
$-1.8\pm1.2\mbox{\,mas\,yr}^{-1}$ in right ascension and
$-9.3\pm5\mbox{\,mas\,yr}^{-1}$ in declination. At $d=1.7$\,kpc
(\cite{prc95}), the difference corresponds to a pulsar velocity
relative to the cluster of $78\pm40\,\mbox{km\,s}^{-1}$---far above the
cluster escape velocity.  Although such a velocity could be imparted
to the system either during the formation of the triple or in an
interaction with another cluster star, the probability of discovering
such a system on an escape trajectory is negligible, since it would
spend only $\sim10^4$\,yr as close to the cluster core as \psr\ is now
observed.  More likely, either the pulsar or cluster proper motion is
in error. Further timing observations are needed to test this claim.
(A chance angular coincidence of an unassociated pulsar and cluster
can be ruled out by the projected position very near the cluster core,
the similar distance estimates, and the relatively good proper motion
agreement.)

The observed rate of change of the dispersion measure,
$(dDM/dt)/DM\approx8\times10^{-6}\mbox{yr}^{-1}$, is comparable to the
$1.4\times10^{-5}\mbox{yr}^{-1}$ change observed in the millisecond
pulsar PSR\,B1937+21, which is at a similar dispersion measure
(\cite{ryb91}).

The only significant deviation from simple Keplerian orbital motion is
a very large derivative of the projected semi-major axis $x$, with
timescale $x_{1a}/\dot x_{1a}\approx3$\,Myr.  The measurement is significant at
$>20\sigma$. Only lower limits are available for the timescales of
changes in the other elements, and the limits are relatively weak in
comparison with the measured timescale of change in $x_{a1}$: $P_a/\dot
P_a\gsim2.2$\,Myr, $e_a/\dot e_a\gsim0.6$\,Myr, and
$|1\,\mbox{rad}/\dot\omega_{1a}|\gsim0.4$\,Myr. We defer discussion of the
orbital perturbations until \S\ref{sec:pert}.

We wish to relate the observed frequency derivatives to the properties 
of the triple system.  To a good first approximation, we can treat the
motion in the system as the sum of two non-interacting Keplerian
ellipses. The pulsar (with mass $m_1$) orbits with the inner
companion (mass $m_2$) around their common center of mass. The
elements of the pulsar's orbit are $x_{1a}$, $P_a$, $e_a$,
$\omega_{1a}$, and $T_a$. The
inner binary (mass $m_1+m_2$) then orbits with the outer companion
(mass $m_3$) around the center of mass of the triple. The
elements of the pulsar's motion in the second orbit are $x_{1b}$,
$P_b$, $e_b$, $\omega_{1b}$, and $T_b$. These latter parameters can be 
related to the line-of-sight acceleration $\bf a\cdot\hat{n}$ where
$\bf \hat{n}$ is a unit vector along the line of sight, which can in
turn be related to the measured frequency derivatives:
\begin{eqnarray}
\dot f &=& -f\frac{\bf a\cdot\hat{n}}{c},\nonumber\\
\ddot f &=&-f\frac{\bf \dot a\cdot\hat{n}}{c},\nonumber\\
&\vdots& \label{eqn:accel}\\
f^{(n)} &=&-f\frac{{\bf a}^{(n-1)}\bf\cdot\hat{n}}{c}\nonumber \\
&\vdots&\nonumber
\end{eqnarray}
where in each case we have neglected terms of order $v/c$ smaller than
the leading contribution.  (As will be seen below, the reflex motion
of the binary in its orbit with the third companion has an amplitude
$\sim6\mbox{\,m\,s}^{-1}\sim2\times10^{-8}c$.) An explicit expansion
of ${\bf a}={\bf a}(x_{1b},P_b,e_b,\omega_{1b},T_b)$ is tedious, but
has been carried out in some special cases by Joshi and Rasio
(1997)\nocite{jr97}. Substitution of the result into
eqn.\,\ref{eqn:accel} yields a series of non-linear equations in five
unknown parameters. Measurement of the five frequency
derivatives $\dot f$,\ldots,$f^{(5)}$ then gives a set of equations
that can be inverted (at least numerically) to determine the orbital
elements. 

In fact, only $\dot f$,\ldots,$f^{(4)}$ have so far been measured in
the case of \psr, with only a limit for $f^{(5)}$
(Table\,\ref{tab:poly}). Hence inversion of eqn.\,\ref{eqn:accel} will
yield a one parameter family of solutions. The situation is further
complicated by the fact that $\dot f$ includes an unknown contribution
from the intrinsic pulsar spin-down rate, so $\dot f=\dot f_{\rm
  int}+\dot f_{\rm acc}$. Because $\dot f_{\rm int}$ is nearly
constant, while $\dot f$ has changed by a factor of five during the time
period described, it is likely that $\dot f\sim\dot f_{\rm acc}$. It
is, on the other hand, unlikely that $\dot f\sim\dot f_{\rm int}$,
since such a large intrinsic torque implies a dipole field strength
$\sim3\times10^9$\,G, an order of magnitude larger than the typical
field for millisecond pulsars in the plane of the Galaxy
(\cite{ctk94}).  Although we recognize the possibility that fields of
pulsars in the plane and in clusters might differ systematically, we assume
a magnetic field strength of
$3\times10^8$\,G, implying that $\dot f_{\rm int}\sim10^{-2}\dot f$, or
$\dot f\approx\dot f_{\rm acc}$. Intrinsic contributions to higher
order derivatives of $f$ are expected to be negligible.

Using the method outlined by Joshi and Rasio (1997), we have inverted
eqn.\,\ref{eqn:accel}, letting the eccentricity $e_b$ range over
discrete values between 0 and 1. For each value of $e_b$, we have used 
the resulting orbital parameters to calculate $f^{(5)}$, and compared
it with the measured limit.  The results are shown in
Fig.\,\ref{fig:polyresults}.
Orbits with eccentricity below $e_b=0.17$ are excluded because they
predict a larger $f^{(5)}$ than observed.  To test the sensitivity of
our results to the assumption $\dot f_{\rm acc}=\dot f_{\rm obs}$, we
have repeated the calculation assuming $\dot f_{\rm acc}=0.1\dot
f_{\rm obs}$. As seen in the figure, relaxing this assumption makes
virtually no change in the inferred properties of the companion or its
orbit (except reducing the mass $m_3$ by about 10\%).  However, orbits
with smaller eccentricity (including circular orbits) are allowed.  We
will return to this point below.

\centerline{\epsfig{file=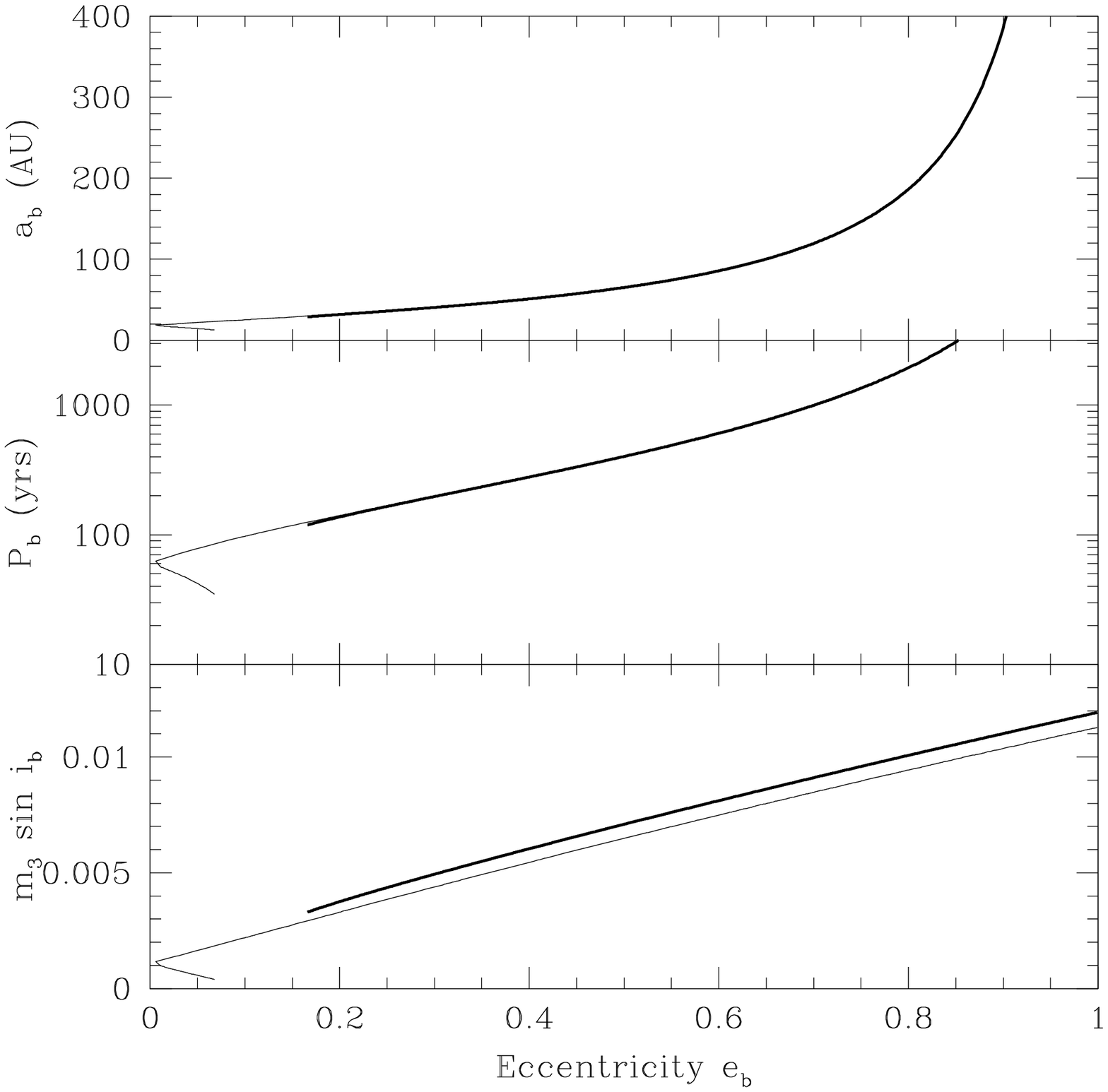,width=0.9\columnwidth}}
\figcaption{\label{fig:polyresults}The one parameter family of solutions
  for the orbit of the outer companion, using the technique of Joshi
  and Rasio (1997).  The inner binary mass was assumed to be
  $m_1+m_2=1.7M_\odot$. The bold line is the solution assuming that
  the observed $\dot f=-5.5\times10^{-15}\,\mbox{s}^{-2}$ on
  JD2448725.5 has negligible contribution from the intrinsic
  spin-down. The lighter line assumes that 90\% of the spin-down rate
  on that date was intrinsic. These curves bound the likely region.}

\subsection{Double Keplerian Models}

Within a simple Keplerian model, there are only five orbital elements
accessible from line-of-sight velocity measurements, so only five
coefficients in the polynomial expansion are independent.
Continuation of the expansion beyond five terms could thus, in
principle, provide a test of the triple hypothesis. Unfortunately, the
current data span is inadequate to provide high-significance estimates
of the terms beyond $f^{(5)}$ in the expansion. Even with the current
data set, a polynomial model is not the optimal approach to estimating
orbital parameters.  For example, high order polynomial terms are
significantly covariant (particularly odd terms with odd, and even
terms with even).  Also, the non-uniformity of the data, both in
sampling intervals and signal-to-noise ratio achieved, makes it
difficult to interpret the fit over the extended data set as a simple
Taylor expansion around a particular epoch.  To avoid these
complexities, we introduce a timing model in which the pulsar orbits
in a hierarchical triple system: it is assumed to move in a Keplerian
orbit around its common center of mass with the inner companion
(possibly with elements that vary linearly with time), and the inner
binary then orbits about the center of mass of the triple in a second
Keplerian elipse.  Not surprisingly, the second orbit is
underdetermined (ignoring for the moment the evolution of the elements
of the inner orbit). We approach this limitation in several ways.

\begin{deluxetable}{ll}
\tablecolumns{2}
\tablewidth{0pc}
\tablecaption{\label{tab:circ}Timing solution for circular outer 
orbit\tablenotemark{a}}
\tablehead{\colhead{Timing parameter}&\colhead{Value (error)}}
\startdata
Spin period $P$ (ms) & 11.075750687(5) \\
Spin frequency $f$ (Hz) & 90.28733386(4) \\
$\dot f$ (s$^{-2}$) & $-4.836(10)\times10^{-15}$ \\
Epoch of $f$ (JD) & 2448725.5 \\
\\
Projected semi-major axis $x$ (s) & 6.4(2) \\
Orbital period $P_b$ (yrs) & 61.8(7) \\
Time of ascending node $T_0$ (JD) & 2449104(5) \\
Mass function ($M_\odot$) & $5.6(4)\times 10^{-10}$ \\
\enddata
\tablenotetext{a}{Intrinsic spin frequency derivatives beyond the first are
  assumed to vanish. Shown are spin parameters and parameters of the
  outer orbit; other parameters are consistent with those given in 
  Table\,\ref{tab:poly}.}
\end{deluxetable}

First, we assume the outer orbit is circular.  The resulting orbital
parameters are given in Table\,\ref{tab:circ}.
If we assume that the mass of the inner binary is $1.7M_\odot$, then
the mass function gives a ``projected'' mass of the second companion
$m_3\sin i=1.2\times10^{-3}M_\odot$ (about 20\% more massive than
Jupiter) and a separation of 19\,AU (comparable to the size of the
orbit of Uranus).  Comparing Tables\,\ref{tab:poly}
and\,\ref{tab:circ}, we note that in the model with a circular outer
orbit, acceleration contributes just over 10\% of the observed
spin-down at the epoch. Our result can thus be directly compared to
the lighter curve in Fig.\,\ref{fig:polyresults}. We note the
excellent agreement (within the published errors) between predictions
from the polynomial model and the direct fit to the double-Keplerian
orbit.

The circular orbit solution yields a relatively large intrinsic
spin-down rate, and hence a large inferred dipole field strength
of $2.6\times10^{9}$\,G.  As discussed above, this is relatively high
for a millisecond pulsar, though well below the $\sim8\times10^9$\,G
upper limit for an 11\,ms pulsar on the spin-up line
(\cite{ls98}). The age of the pulsar, assuming it was born on the
spin-up line and slowed with a braking index $n=3$, is about
$240$\,Myr---far smaller than the cluster age.  If the neutron star
was formed in the collapse of a massive star during the early life of
the cluster, its spin-up to form a millisecond pulsar would be a
relatively recent event.  At first glance, it might be surprising to
note that in this model the ascending node passage occurred during our 
limited data span, in April 1993.  However, with eleven years of data
and a 62\,yr orbit the chances were actually better than one in three
that our observations would include either an ascending or descending
node passage.

\begin{deluxetable}{ll}
\tablecolumns{2}
\tablewidth{0pc}
\tablecaption{\label{tab:ecc}Representative solutions for elliptical outer 
orbit}
\tablehead{\colhead{Binary parameter} & \colhead{Value (error)}}
\startdata
\cutinhead{Eccentricity $e=0.20$}
Projected semi-major axis $x$ (s) & 30.4(1.1) \nl
Orbital period $P_b$ (yrs) & 129(2) \nl
Argument of periastron $(^\circ)$ & $283.9(9)$ \nl
Epoch of periastron $T_0$ (JD) & 2445156(12) \nl
Mass function ($M_\odot$) & $1.36(10)\times 10^{-8}$ \nl
``Projected mass\tablenotemark{a}'' $m_3\sin i_b$ ($M_\odot$) & 
$3.4(1)\times10^{-3}$ \nl
Relative semimajor axis\tablenotemark{b}\, $a_b$ (AU) & 30(2) \nl
\cutinhead{Eccentricity $e=0.50$}
Projected semi-major axis $x$ (s) & 126(4) \nl
Orbital period $P_b$ (yrs) & 389(5) \nl
Argument of periastron $(^\circ)$ & 313.4(5) \nl
Epoch of periastron $T_0$ (JD) & 2446624(10) \nl
Mass function ($M_\odot$) & $1.06(8)\times 10^{-7}$ \nl
``Projected mass\tablenotemark{a}'' $m_3\sin i_b$ ($M_\odot$) & 
$6.7(2)\times10^{-3}$ \nl
Relative semimajor axis\tablenotemark{b}\, $a_b$ (AU) & 64(3) \nl
\enddata
\tablenotetext{a}{Assuming inner binary mass $m_1+m_2=1.7M_\odot$.}
\tablenotetext{b}{The semimajor axis of the relative orbit, 
  $a_b=a_{1b}+a_{2b}$, is nearly independent of $\sin i_b$.}
\end{deluxetable}

As an alternate approach to the underdetermination of the outer
orbital parameters, we could assume that the magnetic field of the
pulsar is small (e.g., $\sim3\times10^8$, a typical value for fast
pulsars (\cite{ctk94})). Then the intrinsic $\dot f$ contributes
negligibly to the observed value, so we take it to be zero. As with
the polynomial fits described above, we are then able to produce a one 
parameter family of solutions, where it is convenient to take the free 
parameter as the eccentricity.  We present orbital parameters for two
values of $e$ in Table\,\ref{tab:ecc}.
In each case, the results from the timing fits agree to better than
10\% (and generally within 5\%) with the parameters obtained from the
polynomial fitting procedure in \S\ref{sec:poly}.

\subsection{Orbital perturbations}\label{sec:pert}

To this point, we have neglected three-body effects. In a hierarchical
triple like the \psr\ system, the orbits are at all times very nearly
Keplerian ellipses. It is customary to consider the osculating orbital
elements which are, at any moment, the Keplerian elements of the orbit
tangent to the real orbit with the same velocity. Various algebraic
and numeric techniques have been developed to determine the time
evolution of these osculating elements (e.g., \cite{bc61}).  We note
that variations in the projected orbital elements can also arise from
proper motion, but such effects are expected to be small in this case
(\cite{ajrt96}). 

The perturbations can be usefully divided into short-period ($\sim
P_a$) terms, long-period ($\sim P_b$) terms, and ``apse-node'' terms
($\sim P_b^2/P_a$) terms (\cite{bro36b,sod75}).  The amplitudes of the
short-period terms are too small to be detected in the current data.
Because the data span is much shorter than either the long-period or
apse-node timescales, the observed ``secular'' perturbations are a sum
of contributions from both terms.  The general solution is quite
complex, but can be greatly simplified if we assume that $P_b$ is much
longer than the data span, so we can take the companion to be at a
fixed position during the observations.  Given the relatively crude
perturbation measurements now available, this assumption is adequate
for even the smallest allowed values of $P_b$.

The orbital perturbations were first calculated in this approximation
by Rasio (1994).  They can be written
(\cite{ras94,jr97})
\begin{eqnarray}
\dot\omega_{1a}&=&\frac{3\pi\eta}{P_a}\left[\sin^2\theta_3\left(5\cos^2\phi_3-1\right)-1\right]\label{eqn:perb1}\\
\dot e_a&=&-\frac{15\pi\eta}{2P_a}e_a\sin^2\theta_3\sin2\phi_3\\
\dot i_a&=&\frac{3\pi\eta}{2P_a}\sin2\theta_3\cos\left(\omega_{1a}+\phi_3\right)\label{eqn:perb3}
\end{eqnarray}
where $\eta=[m_3/(m_1+m_2)](a_a/r_3)^3$, $a_a=a_{1a}+a_{2a}$ is
the semimajor axis of the inner binary, and $r_3$, $\theta_3$, and
$\phi_3$ are the fixed spherical polar coordinates of the second
companion with respect to the center of mass of the inner binary,
choosing $\phi_3$ measured from pericenter in the orbital plane and
$\theta_3$ such that $\sin\theta_3=1$ in the coplanar case. The
``secular'' perturbation of the projected semimajor axis is thus $\dot 
x_{1a}=x_{1a}\cot i_a\dot i_a$.

To use our measurements $\dot x_{1a}$ and constraints on
$\dot\omega_{1a}$ and $\dot e_a$ to further constrain the triple
system parameters, we have performed Monte Carlo simulations following 
the procedure outlined in Joshi and Rasio (1997).  We have assumed a
uniform prior probability distribution for $\cos i_a$, $\cos i_b$, and 
$\alpha$ --- the angle between the lines of nodes of the two
orbits. Whereas Joshi and Rasio assumed a thermal distribution in
$e_b$ (prior probability proportional to $e_b$), without any reason to 
expect that the system is in thermal equilibrium with the cluster, we
have instead adopted a uniform distribution in $e_b$, though we note
that this choice has very little effect on our results.

Our procedure is straightforward.  For each trial, we select values
for $\cos i_a$, $\cos i_b$, $\alpha$, and $e_b$ as described.  Using
$e_b$, we calculate $x_{1b}$, $P_b$, $\omega_{1b}$, and $T_b$ as
described in \S\ref{sec:poly} (assuming $\dot f\approx\dot f_{acc}$).
Using $\cos i_a$, we find $m_2$, and using $\cos i_b$ and $\alpha$ we
calculate $r_3$, $\theta_3$, $\phi_3$, and $\eta$.  (We assume a
neutron star mass $m_1=1.4M_\odot$.)  Using
eqns.~\ref{eqn:perb1}--\ref{eqn:perb3}, we calculate the expected
orbital perturbations, which we compare with the measured values of
$\dot x_{1a}$, $\dot\omega_{1a}$, and $\dot e_a$ and their
uncertainties, rejecting trials with a probability determined from the
appropriate three-dimensional gaussian distribution.

\centerline{\epsfig{file=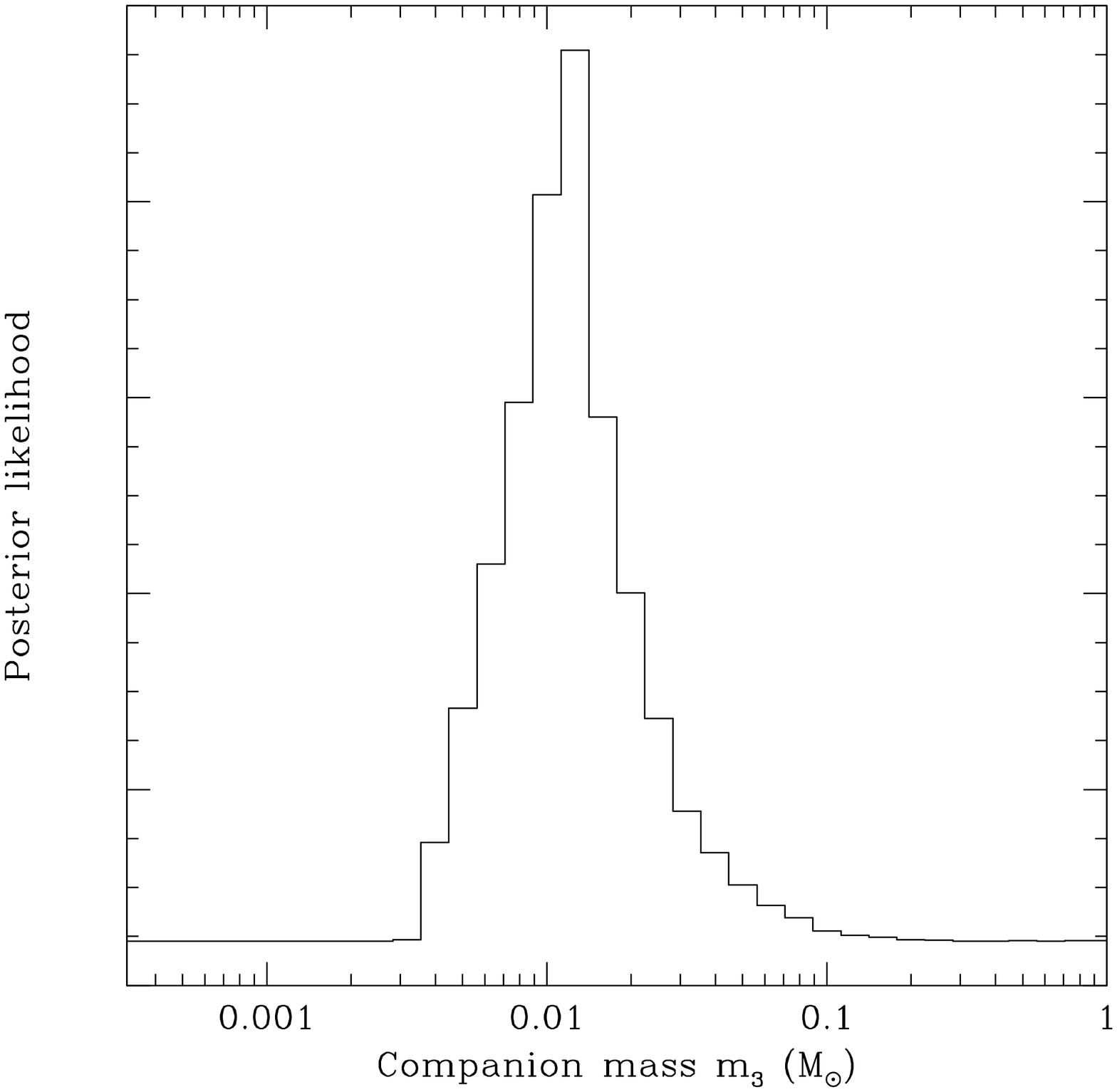,width=0.85\columnwidth}}
\figcaption{\label{fig:m3}Results of Monte Carlo estimation of the
  posterior likelihood distribution for the mass $m_3$ of the outer
  body, as discussed in the text.
}

\centerline{\epsfig{file=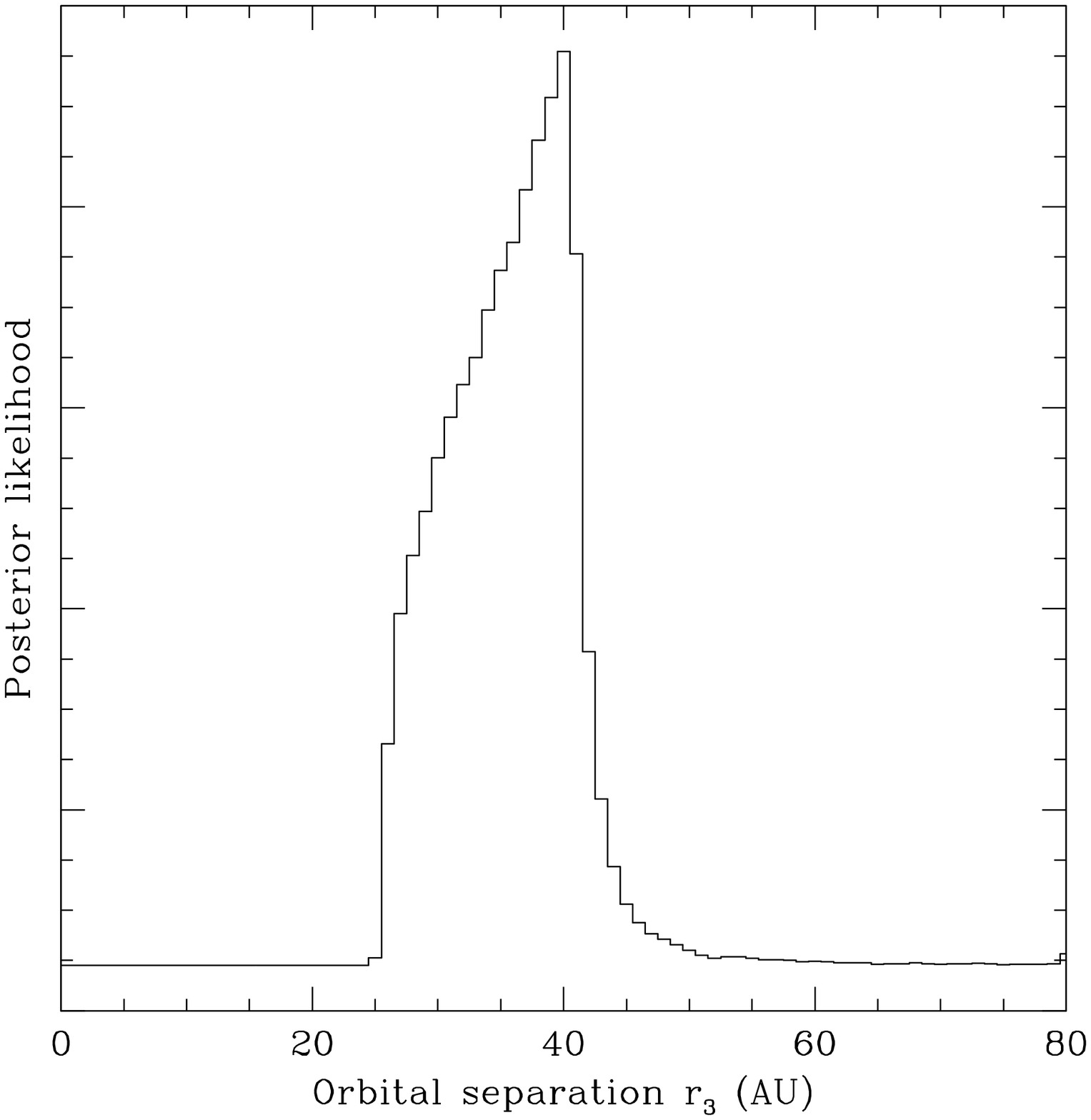,width=0.9\columnwidth}}
\figcaption{\label{fig:r}Results of Monte Carlo estimation of the
  posterior likelihood distribution for the current distance of the
  outer companion from the center of mass of the inner binary, as
  discussed in the text.}
\medskip

Some results of our analysis are shown in
Figs.~\ref{fig:m3}--\ref{fig:incl}.  Our results are in good agreement
with the earlier analysis of Joshi and Rasio (1997), which were based
on a preliminary version of the timing results published here.  We
find that the mass of the outer companion is quite tightly
constrained: $m_3=0.0118^{+0.0087}_{-0.0048}$ (68\%) or
$m_3=0.0118^{+0.0373}_{-0.0073}$ (95\%).  The current distance of the
outer companion from the center of mass of the binary is also well
constrained, $r_3=35\pm6$\,AU (68\%) or $r_3=35\pm10$\,AU (95\%).
Less well constrained is the orbital period $P_b$.  The most favored
values are just above the minimum allowed from the $f^{(n)}$
measurements, or a few hundred years, but there is a long tail to very
long orbital periods.  These solutions correspond to very high
eccentricity orbits in which the second companion is currently very
near periastron.  The 68\% confidence upper limit to the orbital
period is 1200\,yrs.  The inclination of the outer orbit is not well
constrained by the data, with a posterior likelihood very close to the
assumed $\cos i$ prior likelihood.  However, the inner inclination is
constrained to $40\pm12^\circ$ (68\%) or $40\pm24^\circ$ (95\%).  This
leads to a most likely inner companion mass of $m_2=0.46M_\odot$,
somewhat higher that the $0.3M_\odot$ normally assumed.

\centerline{\epsfig{file=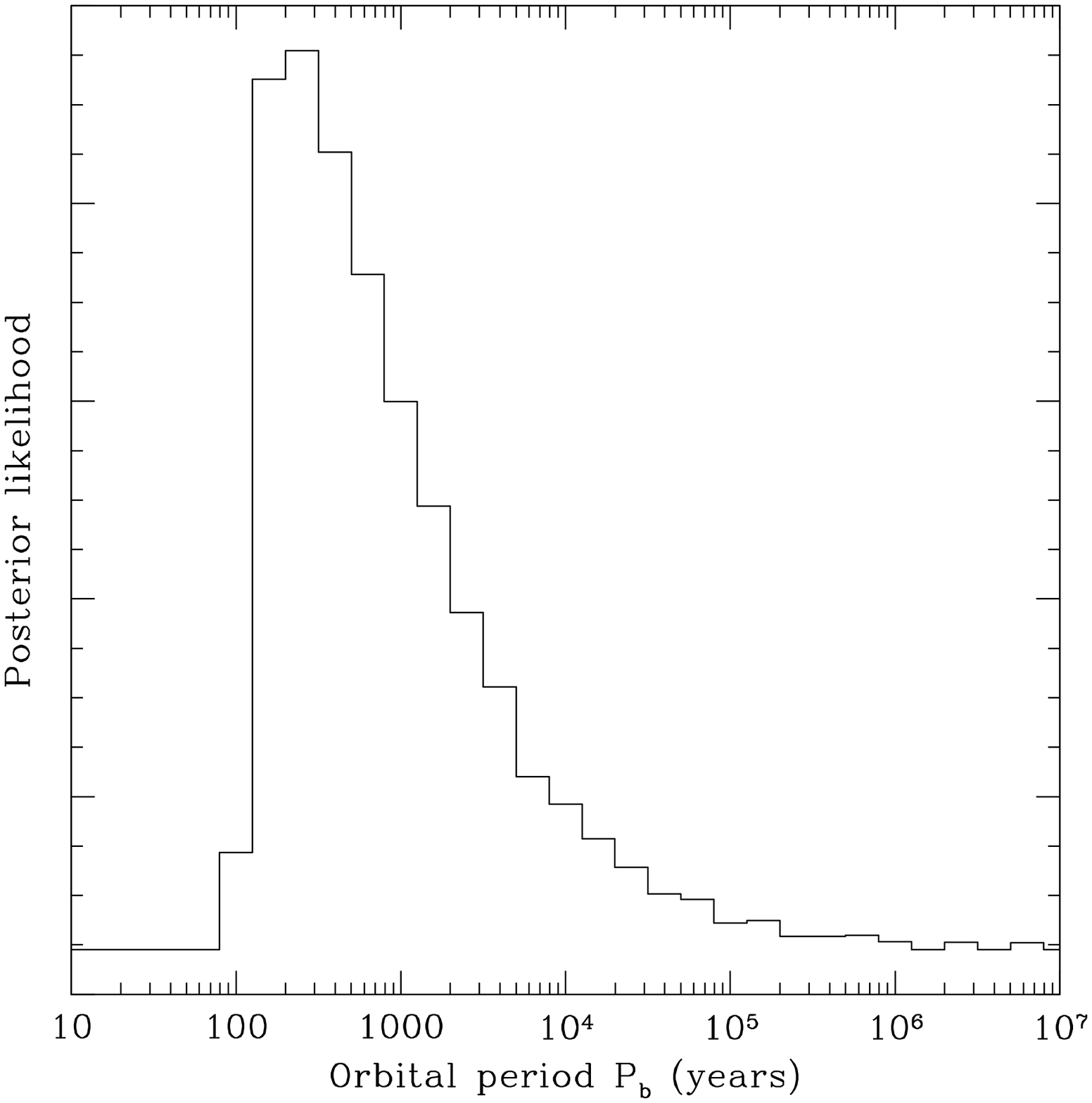,width=0.9\columnwidth}}
\figcaption{\label{fig:pb}Results of Monte Carlo estimation of the
  posterior likelihood distribution for the orbital period of the
  outer orbit, as discussed in the text.}

\centerline{\epsfig{file=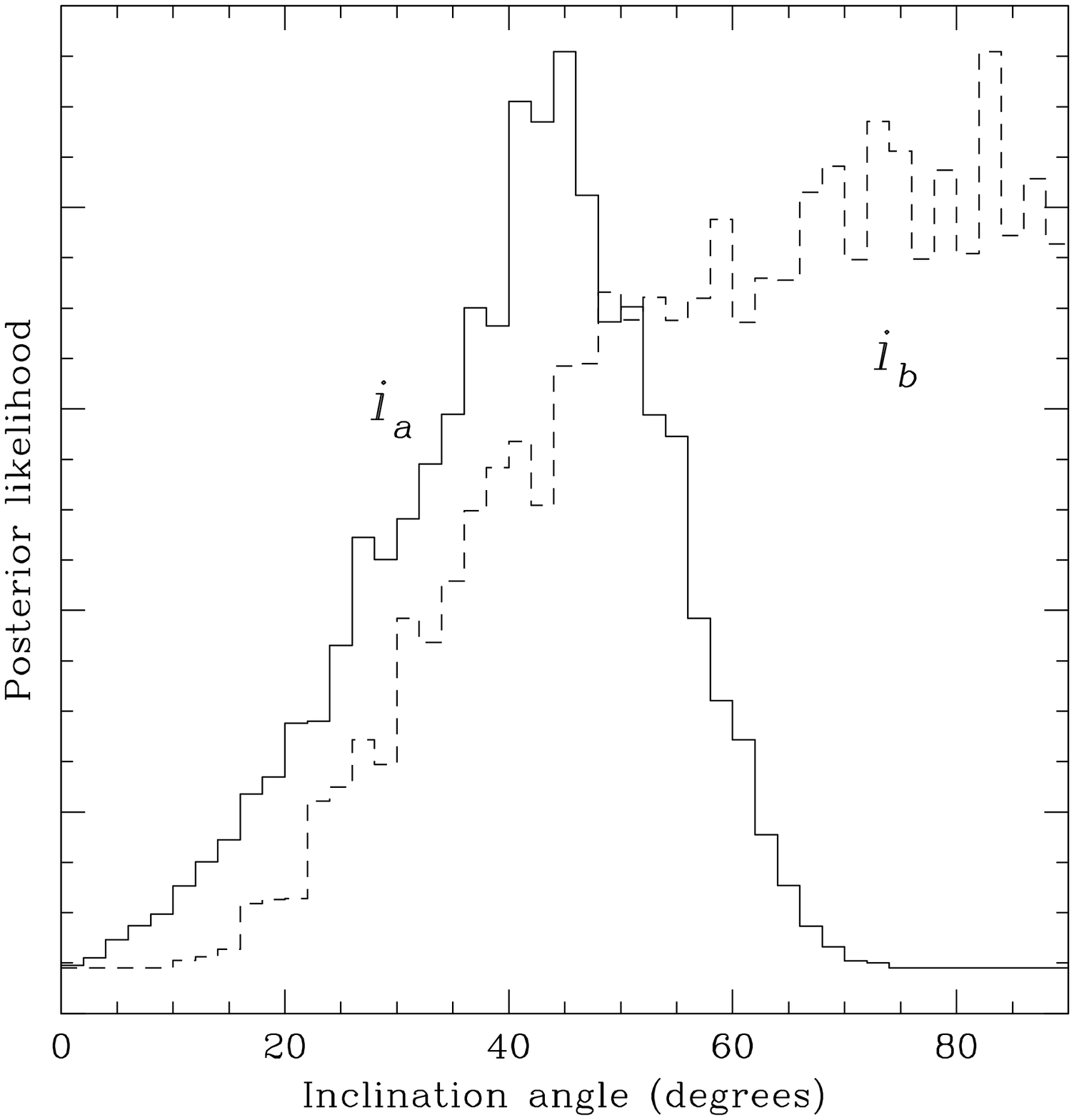,width=0.9\columnwidth}}
\figcaption{\label{fig:incl}Results of Monte Carlo estimation of the
  posterior likelihood distribution for the inclinations of the inner
  and outer ``binary'' orbits, as discussed in the text.  The
  inclination of the outer orbit is not strongly constrained (the
  prior distribution was uniform in $\cos i_b$), but the inner orbit
  is constrained to a fairly low inclination, $i_a\sim40^\circ$.}
\bigskip

We have also confirmed that the orbital perturbation measurements and
limits are consistent with the solution in which $\dot f_{acc}=0.1\dot
f_{obs}$. In particular, solutions with a circular outer orbit
(Table~2) are acceptable if $i_a\sim40^\circ$ and $\alpha\sim100^\circ$ or
$\alpha\sim280^\circ$.  Again, $i_b$ is poorly constrained, and
$m_3=1.55\pm0.30\times10^{-3}M_\odot$, or about half again the mass of 
Jupiter.

\section{Conclusions}\label{sec:discussion}



Since the initial suggestion that pulsar \psr\ is the member of a
triple system, constraints on the properties of the system components
have gradually improved.  The detection of a secular change in the
projected semimajor axis of the inner binary, which has been discussed
previously at conferences (e.g., \cite{ajrt96}) but is presented in
detail here for the first time, is an important confirmation of the
triple nature of the system, since there is no known way for spin
instabilities of the pulsar to produce timing fluctuations on the
orbital timescale.  The only simple way to understand the system is as 
a binary accelerating in an external tidal field, most likely due to a 
bound companion.

As we have shown, the orbital parameters of the outer orbit are still
poorly constrained.  The reason is not hard to understand.  Because
the span of available data is much shorter than the period of the
outer orbit, the relative position of the inner binary and outer
companion has not changed substantially since the pulsar was
discovered. Although the position itself is known rather accurately,
the relative velocity is not.  This leads to a family of allowed
solutions, all with periastron distance $\sim35$\,AU (the current
separation).  Very high eccentricity orbits with very large semimajor
axis can account for the observed acceleration and orbital
perturbations, but only if the system is currently observed near
periastron.  Because a body moving in a high eccentricity orbit spends
only a small fraction of the time near periastron, such solutions
require significant fine tuning.

Another remaining source of uncertainty is the extent to which the
observed frequency derivative $\dot f_{obs}$ is dominated by acceleration,
rather than intrinsic pulsar spin-down.  We have argued that it is
most likely that $\dot f_{acc}\approx\dot f_{obs}$, in which case
$m_3\sim0.01M_\odot$. Allowing a significant contribution from $\dot
f_{int}$ reduces the magnitude of the acceleration by the third body,
allowing masses as small as $m_3\sim10^{-3}M_\odot$.


In either case, the second companion is almost certain substellar:
below the $\sim0.08M_\odot$ hydrogen burning limit, and is most likely
below or near the deuterium burning limit, $\sim0.015M_\odot$
(\cite{lp87b}).  Whether it is called a ``brown dwarf'' or a
``planet'' is probably not important.  The possibility that such
objects might be found in globular clusters was first suggested just
prior to the discovery of the triple nature of \psr\ (\cite{sig92}),
and Sigurdsson (1993, 1995)\nocite{sig93b,sig95} has further suggested
models for the formation of such a triple involving exchange
interactions in the dense environment of M4.  However, developing a
complete model that can explain not only the system formation and
stability and the pulsar spin-up, but also the non-zero eccentricity
of the inner binary remains an open problem (\cite{jr97}).

The prospects for continued improvement of the timing constraints on
the system parameters are good.  We expect
(eqns.~\ref{eqn:perb1}--\ref{eqn:perb3}) that the timescales for
perturbation of $\omega_{1a}$ and $e_a$ should be comparable to that
of $x_{1a}$, and as noted in \S\ref{sec:poly} the current measurement
uncertainties have nearly reached that level.  Furthermore, the limits
on $f^{(5)}$ already significantly constrain the allowed solutions,
and we expect the uncertainty to improve rapidly with observing
span.  

\acknowledgments
We thank D. Backer, R. Foster, and J. Taylor for their substantial
contributions to the early stages of this project, and an anonymous
referee for helpful comments.  M. McKinnon,
T. Hankins, and M. Goss provided important support for pulsar timing
at the VLA.  We particularly thank K. Joshi, F. Rasio, and
S. Sigurdsson for many interesting and helpful discussions over the
years.  Green Bank and the VLA are facilities of the National Radio
Astronomy Observatory, operated by Associated Universities, Inc.,
under contract from the NSF, which has also provided direct support
for this work---as have Caltech, Princeton, and the Sloan Foundation.


\end{document}